\input epsf

\newfam\scrfam
\batchmode\font\tenscr=rsfs10 \errorstopmode
\ifx\tenscr\nullfont
        \message{rsfs script font not available. Replacing with calligraphic.}
        \def\scr{\cal}
\else   
        \font\sevenscr=rsfs7
        \font\fivescr=rsfs5
        \skewchar\tenscr='177 \skewchar\sevenscr='177 \skewchar\fivescr='177
        \textfont\scrfam=\tenscr \scriptfont\scrfam=\sevenscr
        \scriptscriptfont\scrfam=\fivescr
        \def\scr{\fam\scrfam}
        
\fi
\catcode`\@=11
\newfam\frakfam
\batchmode\font\tenfrak=eufm10 \errorstopmode
\ifx\tenfrak\nullfont
        \message{eufm font not available. Replacing with italic.}
        
\else
    
    \font\sevenfrak=eufm7 \font\fivefrak=eufm5
    \textfont\frakfam=\tenfrak
    \scriptfont\frakfam=\sevenfrak \scriptscriptfont\frakfam=\fivefrak
    
\fi
\catcode`\@=\active
\newfam\msbfam
\batchmode\font\twelvemsb=msbm10 scaled\magstep1 \errorstopmode
\ifx\twelvemsb\nullfont\def\Bbb{\bf}

    \message{Blackboard bold not available. Replacing with boldface.}
\else   \catcode`\@=11
        \font\tenmsb=msbm10 \font\sevenmsb=msbm7 \font\fivemsb=msbm5
        \textfont\msbfam=\tenmsb
        \scriptfont\msbfam=\sevenmsb \scriptscriptfont\msbfam=\fivemsb
        \def\Bbb{\relax\expandafter\Bbb@}
        \def\Bbb@#1{{\Bbb@@{#1}}}
        \def\Bbb@@#1{\fam\msbfam\relax#1}
        \catcode`\@=\active

\fi
\newfam\cpfam
\def\sectionfonts{\relax
    \textfont0=\twelvecp          \scriptfont0=\ninecp
      \scriptscriptfont0=\sixrm
    \textfont1=\twelvei           \scriptfont1=\ninei
      \scriptscriptfont1=\sixi
    \textfont2=\twelvesy           \scriptfont2=\ninesy
      \scriptscriptfont2=\sixsy
    \textfont3=\twelveex          \scriptfont3=\tenex
      \scriptscriptfont3=\tenex
    \textfont\itfam=\twelveit     \scriptfont\itfam=\nineit
    \textfont\slfam=\twelvesl     \scriptfont\slfam=\ninesl
    \textfont\bffam=\twelvebf     \scriptfont\bffam=\ninebf
      \scriptscriptfont\bffam=\sixbf
    \textfont\ttfam=\twelvett
    \textfont\cpfam=\twelvecp
}
        \font\eightrm=cmr8              \def\xrm{\eightrm}
        \font\eightbf=cmbx8             \def\xbf{\eightbf}
        \font\eightit=cmti10 at 8pt     \def\xit{\eightit}
                       
        \font\sixrm=cmr6                
                     
        \font\eightcp=cmcsc8
        \font\eighti=cmmi8              \def\xold{\eighti}
        \font\eightib=cmmib8             \def\xbold{\eightib}
        \font\teni=cmmi10               \def\old{\teni}
        \font\ninei=cmmi9
        \font\tencp=cmcsc10
        \font\ninecp=cmcsc9

        \font\twelvei=cmmi12
        \font\twelvecp=cmcsc10 scaled\magstep1

        \font\twelvesy=cmsy12
        \font\ninesy=cmsy9
        \font\sixsy=cmsy6
        \font\twelveex=cmex12

        \font\twelveit=cmti12
        \font\nineit=cmti9
        
        \font\twelvesl=cmsl12
        \font\ninesl=cmsl9
        
        \font\twelvebf=cmbx12
        \font\ninebf=cmbx9
        \font\sixbf=cmbx6
        \font\twelvett=cmtt12

        \font\sixi=cmmi6

\batchmode\font\tenhelvbold=phvb at10pt \errorstopmode
\ifx\tenhelvbold\nullfont
        \message{phvb font not available. Replacing with cmr.}
    \font\tenhelvbold=cmb10   
    \font\twelvehelvbold=cmb12
    
    \font\sixteenhelvbold=cmb16
  \else
    \font\tenhelvbold=phvb at10pt   
    \font\twelvehelvbold=phvb at12pt
     at14pt
    \font\sixteenhelvbold=phvb at16pt
\fi

\def\noblackbox{\overfullrule=0pt}
\noblackbox

\newtoks\headtext
\headline={\ifnum\pageno=1\hfill\else
    \ifodd\pageno{\eightcp\the\headtext}{ }\dotfill{ }{\old\folio}
    \else{\old\folio}{ }\dotfill{ }{\eightcp\the\headtext}\fi
    \fi}
\def\makeheadline{\vbox to 0pt{\vss\noindent\the\headline\break
\hbox to\hsize{\hfill}}
        \vskip2\baselineskip}
\newcount\infootnote
\infootnote=0
\def\foot#1#2{\infootnote=1
\footnote{${}^{#1}$}{\vtop{\baselineskip=.75\baselineskip
\advance\hsize by
-\parindent\noindent{\xrm #2\hfill\vskip\parskip}}}\infootnote=0$\,$}
\newcount\refcount
\refcount=1
\newwrite\refwrite
\def\oldsize{\ifnum\infootnote=1\xold\else\old\fi}
\def\ref#1#2{
    \def#1{{{\oldsize\the\refcount}}\ifnum\the\refcount=1\immediate\openout\refwrite=\jobname.refs\fi\immediate\write\refwrite{\item{[{\xold\the\refcount}]}
    #2\hfill\par\vskip-2pt}\xdef#1{{\noexpand\oldsize\the\refcount}}\global\advance\refcount by 1}
    }
\def\refout{\catcode`\@=11
        \xrm\immediate\closeout\refwrite
        \vskip2\baselineskip
        {\noindent\twelvecp References}\hfill
        \par\nobreak\vskip\baselineskip
        \baselineskip=.75\baselineskip
        \input\jobname.refs
        \baselineskip=4\baselineskip \divide\baselineskip by 3
        \catcode`\@=\active\rm}

\def\hepth#1{\href{http://arxiv.org/abs/hep-th/#1}{arXiv:hep-th/{\xold#1}}}

\def\jhep#1#2#3#4{\href{http://jhep.sissa.it/stdsearch?paper=#2\%28#3\%29#4}{J. High Energy Phys. {\xbold #1#2} ({\xold#3}) {\xold#4}}}

\def\ATMP#1#2#3{Adv. Theor. Math. Phys. {\xbold#1} ({\xold#2}) {\xold#3}}

\def\CQG#1#2#3{Class. Quantum Grav. {\xbold#1} ({\xold#2}) {\xold#3}}

\def\JHEP{\jhep}

\def\NPB#1#2#3{Nucl. Phys. {\xbf B}{\xbold#1} ({\xold#2}) {\xold#3}}

\newcount\sectioncount
\sectioncount=0
\def\section#1#2{\global\eqcount=0
    \global\subsectioncount=0
        \global\advance\sectioncount by 1
    \ifnum\sectioncount>1
            \vskip2\baselineskip
    \fi
    \noindent
       \line{\sectionfonts\twelvecp\the\sectioncount. #2\hfill}
        \par\nobreak\vskip.8\baselineskip\noindent
        \xdef#1{{\old\the\sectioncount}}}
\newcount\subsectioncount
\def\subsection#1#2{\global\advance\subsectioncount by 1
    \par\nobreak\vskip.8\baselineskip\noindent
    \line{\tencp\the\sectioncount.\the\subsectioncount. #2\hfill}
    \vskip.5\baselineskip\noindent
    \xdef#1{{\old\the\sectioncount}.{\old\the\subsectioncount}}}
\newcount\appendixcount
\appendixcount=0
\def\appendix#1{\global\eqcount=0
        \global\advance\appendixcount by 1
        \vskip2\baselineskip\noindent
        \ifnum\the\appendixcount=1
        \hbox{\twelvecp Appendix A: #1\hfill}
        \par\nobreak\vskip\baselineskip\noindent\fi
    \ifnum\the\appendixcount=2
        \hbox{\twelvecp Appendix B: #1\hfill}
        \par\nobreak\vskip\baselineskip\noindent\fi
    \ifnum\the\appendixcount=3
        \hbox{\twelvecp Appendix C: #1\hfill}
        \par\nobreak\vskip\baselineskip\noindent\fi}
\def\acknowledgements{\vskip2\baselineskip\noindent
        \underbar{\it Acknowledgements:}\ }
\newcount\eqcount
\eqcount=0
\def\Eqn#1{\global\advance\eqcount by 1
\ifnum\the\sectioncount=0
    \xdef#1{{\old\the\eqcount}}
    \eqno({\oldstyle\the\eqcount})
\else
        \ifnum\the\appendixcount=0
            \xdef#1{{\old\the\sectioncount}.{\old\the\eqcount}}
                \eqno({\oldstyle\the\sectioncount}.{\oldstyle\the\eqcount})\fi
        \ifnum\the\appendixcount=1
            \xdef#1{{\oldstyle A}.{\old\the\eqcount}}
                \eqno({\oldstyle A}.{\oldstyle\the\eqcount})\fi
        \ifnum\the\appendixcount=2
            \xdef#1{{\oldstyle B}.{\old\the\eqcount}}
                \eqno({\oldstyle B}.{\oldstyle\the\eqcount})\fi
        \ifnum\the\appendixcount=3
            \xdef#1{{\oldstyle C}.{\old\the\eqcount}}
                \eqno({\oldstyle C}.{\oldstyle\the\eqcount})\fi
\fi}
\def\eqn{\global\advance\eqcount by 1
\ifnum\the\sectioncount=0
    \eqno({\oldstyle\the\eqcount})
\else
        \ifnum\the\appendixcount=0
                \eqno({\oldstyle\the\sectioncount}.{\oldstyle\the\eqcount})\fi
        \ifnum\the\appendixcount=1
                \eqno({\oldstyle A}.{\oldstyle\the\eqcount})\fi
        \ifnum\the\appendixcount=2
                \eqno({\oldstyle B}.{\oldstyle\the\eqcount})\fi
        \ifnum\the\appendixcount=3
                \eqno({\oldstyle C}.{\oldstyle\the\eqcount})\fi
\fi}
\def\multi{\global\advance\eqcount by 1}
\def\multieq#1#2{
    \ifnum\the\sectioncount=0
        \eqno({\oldstyle\the\eqcount})
         \xdef#1{{\old\the\eqcount#2}}
    \else
        \xdef#1{{\old\the\sectioncount}.{\old\the\eqcount}#2}
        \eqno{({\oldstyle\the\sectioncount}.{\oldstyle\the\eqcount}#2)}
    \fi}

\newtoks\url
\def\Href#1#2{\catcode`\#=12\url={#1}\catcode`\#=\active#2}
\def\href#1#2{{#2}}

\parskip=3.5pt plus .3pt minus .3pt
\baselineskip=14pt plus .1pt minus .05pt
\lineskip=.5pt plus .05pt minus .05pt
\lineskiplimit=.5pt
\abovedisplayskip=18pt plus 4pt minus 2pt
\belowdisplayskip=\abovedisplayskip
\hsize=14cm
\vsize=19.9cm
\hoffset=1.5cm
\voffset=1.8cm
\frenchspacing
\footline={}
\raggedbottom

\def\ss{\scriptstyle}

\def\*{\partial}
\def\punkt{\,\,.}
\def\komma{\,\,,}

\def\={\!=\!}
\def\small#1{{\hbox{$#1$}}}

\def\fraction#1{\small{1\over#1}}
\def\fr{\fraction}
\def\Fraction#1#2{\small{#1\over#2}}
\def\Fr{\Fraction}

\def\ie{{\tenit i.e.}}

\def\a{\alpha}
\def\b{\beta}

\def\e{\varepsilon}
\def\g{\gamma}

\def\w{\!\wedge\!}

\def\Re{\hbox{Re}\,}




\def\l{\lambda}
\def\th{\theta}

\def\bl{{\l}}
\def\bw{{w}}
\def\bD{{D}}

\def\tw{\tilde w}

\def\PP{{\scr P}}
\def\HH{{\scr H}}

\def\lra{\longrightarrow}

\def\arrowover#1{\vtop{\baselineskip=0pt\lineskip=0pt
      \ialign{\hfill##\hfill\cr$\lra$\cr${\ss #1}$\cr}}}

\def\arrowunder#1{\raise4pt\vtop{\baselineskip=0pt\lineskip=0pt
      \ialign{\hfill##\hfill\cr${\ss #1}$\cr$\lra$\cr}}}

\def\Qarrow{\;\arrowover Q\;}


\ref\CederwallNilssonTsimpisI{M. Cederwall, B.E.W. Nilsson and D. Tsimpis,
{\xit ``The structure of maximally supersymmetric super-Yang--Mills theory---constraining higher order corrections''}, \jhep{01}{06}{2001}{034}
[\hepth{0102009}].}

\ref\CederwallNilssonTsimpisII{M. Cederwall, B.E.W. Nilsson and D. Tsimpis,
{\xit ``D=10 super-Yang--Mills at $\ss O(\a'^2)$''},
\JHEP{01}{07}{2001}{042} [\hepth{0104236}].}

\ref\CGNN{M. Cederwall, U. Gran, M. Nielsen and B.E.W. Nilsson,
{\xit ``Manifestly supersymmetric M-theory''},
\JHEP{00}{10}{2000}{041} [\hepth{0007035}];
{\xit ``Generalised 11-dimensional supergravity''}, \hepth{0010042}.
}

\ref\CGNT{M. Cederwall, U. Gran, B.E.W. Nilsson and D. Tsimpis,
{\xit ``Supersymmetric corrections to eleven-dimen\-sional supergravity''},
\jhep{05}{05}{2005}{052} [\hepth{0409107}].}

\ref\SpinorialCohomology{M. Cederwall, B.E.W. Nilsson and D. Tsimpis,
{\xit ``Spinorial cohomology and maximally supersymmetric theories''},
\jhep{02}{02}{2002}{009} [\hepth{0110069}];
M. Cederwall, {\xit ``Superspace methods in string theory, supergravity and gauge theory''}, Lectures at the XXXVII Winter School in Theoretical Physics ``New Developments in Fundamental Interactions Theories'',  Karpacz, Poland,  Feb. 6-15, 2001, \hepth{0105176}.}

\ref\SuperYM{L. Brink, J.H. Schwarz and J. Scherk,
{\xit ``Supersymmetric Yang--Mills theories''},
\NPB{121}{1977}{77}.}

\ref\NilssonSixDSYM{B.E.W. Nilsson,
{\xit ``Superspace action for a 6-dimensional non-extended supersymmetric
Yang--Mills theory''},
\NPB{174}{1980}{335}.}

\ref\NilssonSYM{B.E.W.~Nilsson,
\xit ``Off-shell fields for the 10-dimensional supersymmetric
Yang--Mills theory'', \xrm G\"oteborg-ITP-{\xold81}-{\xold6}.}

\ref\NilssonPure{B.E.W.~Nilsson,
{\xit ``Pure spinors as auxiliary fields in the ten-dimensional
supersymmetric Yang--Mills theory''},
\CQG3{1986}{{\xrm L}41}.}

\ref\BerkovitsI{N. Berkovits,
{\xit ``Super-Poincar\'e covariant quantization of the superstring''},
\jhep{00}{04}{2000}{018} [\hepth{0001035}].}

\ref\BerkovitsReview{N. Berkovits, {\xit ``ICTP lectures on covariant
quantization of the superstring''}, \hepth{0209059}.}

\ref\BerkovitsParticle{N. Berkovits, {\xit ``Covariant quantization of
the superparticle using pure spinors''}, \jhep{01}{09}{2001}{016}
[\hepth{0105050}].}

\ref\BerkovitsCohomology{N. Berkovits,
{\xit ``Cohomology in the pure spinor formalism for the superstring''},
\jhep{00}{09}{2000}{046} [\hepth{0006003}].}

\ref\Movshev{M. Movshev and A. Schwarz, {\xit ``On maximally
supersymmetric Yang--Mills theories''}, \NPB{681}{2004}{324}
[\hepth{0311132}].}

\ref\Berkovits{N. Berkovits,
{\xit ``Pure spinor formalism as an N=2 topological string''},
\jhep{05}{10}{2005}{089} [\hepth{0509120}].}

\ref\NiclasW{P. Grassi and N. Wyllard, {\xit ``Lower-dimensional
pure-spinor superstrings''}, \jhep{05}{12}{2005}{007}
[\hepth{0509140}];
N. Wyllard, {\xit ``Pure-spinor superstrings in d=2,4,6''},
\jhep{05}{11}{009}{2005} [\hepth{0509165}].}

\ref\CGN{M. Cederwall, U. Gran and B.E.W. Nilsson, {\xit work in
progress}.}

\ref\HoweTsimpis{P.S. Howe and D. Tsimpis, {\xit ``On higher-order
corrections in M theory''}, \jhep{03}{09}{2003}{038}
[\hepth{0305129}].}

\ref\GrassiVanhove{P.A. Grassi and P. Vanhove, {\xit ``Topological M
theory from pure spinor formalism''}, \ATMP{9}{2005}{285}
[\hepth{0411167}].}

\ref\Koller{J. Koller, {\xit ``A six-dimensional superspace approach
to extended superfields''}, \NPB{222}{1983}{319}.}

\ref\HoweSierraTownsend{P.S. Howe, G. Sierra and P.K. Townsend,
{\xit ``Supersymmetry in six dimensions''}, \NPB{221}{1983}{331}.}

\ref\BSvP{E. Bergshoeff, E. Sezgin and A. van Proeyen,
{\xit ``Superconformal tensor calculus and matter couplings
in six dimensions''}, \NPB{264}{1986}{653}.}


\headtext={Cederwall, Nilsson: ``Pure spinors and D=6 super-Yang--Mills''}

\line{
\epsfysize=15mm
\epsffile{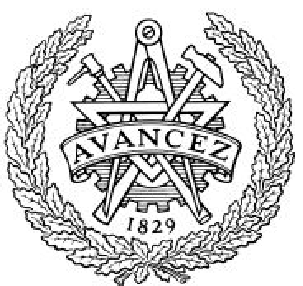}
\hfill}
\vskip-12mm
\line{\hfill G\"oteborg preprint}
\line{\hfill January, {\old2008}}
\line{\hrulefill}

\vfill
\vskip.5cm

\centerline{\sixteenhelvbold
Pure Spinors and D=6 Super-Yang--Mills}

\vfill

\centerline{\twelvehelvbold
Martin Cederwall and Bengt EW Nilsson}

\vfill

\centerline{\it Fundamental Physics}
\centerline{\it Chalmers University of Technology}
\centerline{\it SE 412 96 G\"oteborg, Sweden}

\vfill

{\narrower\noindent \underbar{Abstract:} Pure spinor cohomology has
been used to describe maximally supersymmetric theories, like $D=10$
super-Yang--Mills and $D=11$ supergravity. Supersymmetry closes
on-shell in such theories, and the fields in the cohomology
automatically satisfy the equations of motion. In this paper, we
investigate the corresponding structure in a model with {\it off-shell}
supersymmetry, $N=1$ super-Yang--Mills theory in
$D=6$. Here, fields and antifields are obtained as cohomologies in
different complexes with respect to the BRST operator $Q$. It turns
out to be natural to enlarge the pure spinor space with additional
bosonic variables, subject to some constraints generalising the pure
spinor condition, in order to accommodate the different
supermultiplets in the same generalised pure spinor wave-function.
We construct another BRST operator, $s$, acting in
the cohomology of $Q$, whose cohomology implies the equations of motion.
We comment on the possible use of similar approaches in other models.
\smallskip}
\vfill

\font\xxtt=cmtt6

\vtop{\baselineskip=.6\baselineskip\xxtt
\line{\hrulefill}
\catcode`\@=11
\line{email: martin.cederwall@chalmers.se,
tfebn@fy.chalmers.se\hfill}
\catcode`\@=\active
}

\eject

\section\Introduction{Introduction}Pure spinors and pure spinor
cohomology (or spinorial cohomology) seems to be a deep structure
underlying supersymmetric gauge theories, including supergravity.
In a more pragmatic sense, pure spinors act as a book-keeping device for
superspace forms with spinorial indices. The pure spinor constraint,
which is generically of the form $(\l\g^a\l)=0$, essentially projects
out torsion from the anti-commutator of two fermionic derivatives, and
ensures the nilpotency of the pure spinor BRST operator
$Q=\l^\a D_\a$.

Finding superspace formulations of the component dynamics of
supersymmetric Yang--Mills theory [\SuperYM] has to a large extent
been a question of trial and error. It was recognised early that
pure spinors has a r\^ole to play [\NilssonSYM,\NilssonPure], but it
took much longer to realise how to make systematic use of them. Pure
spinor techniques arise naturally from the formulation of gauge theory
on superspace, and have been used successfully in $D=10$ 
[\CederwallNilssonTsimpisI,\BerkovitsParticle,\SpinorialCohomology,\Movshev],
both for the undeformed theory and its supersymmetric deformations.
The theory may be supplemented with extra 
pure spinor variables that enable the construction of a Chern--Simons
like off-shell action [\Berkovits] constructed in terms of $Q$.
These techniques have also been very powerful in the covariant
quantisation of the superstring (see refs.
[\BerkovitsI,\BerkovitsReview] and references therein). Some results
have been obtained for $D=11$ supergravity [\CGNN], and for
maximally supersymmetric models, where explicit higher-derivative
corrections were derived in refs.
[\CederwallNilssonTsimpisI,\CederwallNilssonTsimpisII,\CGNN,\CGNT].

Standard component formulations of maximally supersymmetric theories
are on-shell supersymmetric---the equations of motion are implied by
$Q=0$---whereas the pure spinor framework provides an off-shell
formulation. In this sense, the property of the supermultiplets of
having supersymmetry transformations closing only on shell is turned
into a virtue. The dynamics can be given the simple form of
vanishing curvature (the equations of motion of a Chern--Simons-like
action). The corresponding situation for less supersymmetric
theories has not been as clear. Attempts have been made to use pure
spinors in lower dimensions [\NiclasW], but mainly in the context of
superstrings. Neither has it been as thoroughly investigated, since
the existence of off-shell supermultiplets provides means of
treating the models without sacrificing supersymmetry. Nevertheless,
pure spinor cohomology works as a generic method of defining the
supermultiplets, including auxiliary fields, and if we think there
is some deeper significance to such a statement, we should also
seriously consider the mechanisms used to go from the off-shell
multiplets to the on-shell theory. This question is addressed and
solved in the present paper for a specific case of a theory with 8
supercharges, $N=1$ super-Yang--Mills theory in $D=6$. (The
superspace formulation of this theory, as well of the one in $D=10$
is of course well known
[\NilssonSixDSYM,\HoweSierraTownsend,\Koller,\BSvP].)

In the concluding section we also comment on how similar techniques
may clarify the situation for supermultiplets that are partially
off-shell, like $D=10$, $N=1$ supergravity and heterotic supergravity,
and how they may be used to address questions in $D=11$ supergravity
and M-theory.


\section\DIsTen{Review of pure spinors in $D=10$ and
super-Yang--Mills}Consider
a scalar field, or first-quantised wave function, that in
addition to the superspace coordinates $x^a$, $\theta^\a$ (we take
superspace to be flat, for simplicity) depend on a pure spinor
$\l^\a$. The pure spinor constraint is
$T^a\equiv(\l\g^a\l)\approx0$. The fields is seen as a formal series
expansion in $\l$, and gauge invariance (with the pure spinor
constraint as generator) is implemented by factoring out the ideal
generated by $T$.

The flat superspace torsion is $T^a_{\a\b}=2\g^a_{\a\b}$. The pure
spinor constraint implies that the fermionic operator
$Q=\l^\a D_\a$ is nilpotent, and can be used as a BRST operator.
The content of a scalar field, \ie, the
structure of the complex is
$$
r_{0}\Qarrow r_{1}\Qarrow r_{2}\Qarrow\ldots\Qarrow r_{n}\Qarrow\ldots
\Eqn\BasicComplex
$$
where the representations $r_{n}$ at each $n$ denotes a superfield in
the representation $r_{n}$ of the Lorentz group. This representation
is the conjugate representation to the symmetric product of $n$ pure
spinors.
When we describe
a gauge theory, $r_{n}$ consists of
totally symmetric and $\g$-traceless
tensors in $n$ spinor indices. The fermionic
exterior derivative is a projection on the representations $r_{n}$ of
a (symmetrised) spinorial covariant derivative. The general
interpretation is that $r_{0}$ contains gauge transformations (or
ghosts), $r_{1}$
contains fields, $r_{2}$ deformations (or antifields) and $r_3$
antighosts of the theory.
For a theory of a rank-$(p+1)$ tensor potential, such as the three-form
in $D=11$ supergravity, $r_{p}$ contains gauge transformations
(ghosts), $r_{<p}$ ghosts for ghosts, $r_{p+1}$
fields, $r_{p+2}$ deformations (antifields), $r_{p+3}$ antighosts,
etc. Although the structure seems general, it really occurs only for a
limited number of cases, and relies on maximal supersymmetry.

\epsfxsize=3cm
\vskip2\parskip
\hskip8\parskip\vbox{\hsize=12cm
\epsffile{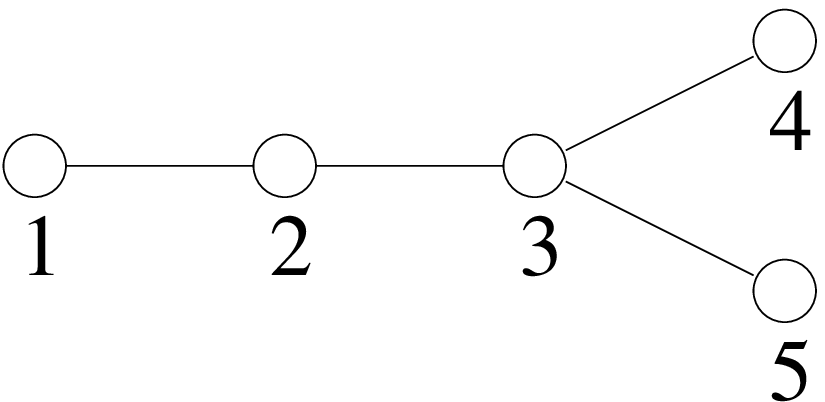}\hfill\break
{\it Figure 1: Convention for Dynkin indices for $so(1,9)$ representations}
}
\vskip2\parskip
For $N=1$ and $D=10$, with $\l^\a$ in $(00001)$ and $D_\a$ in
$(00010)$, we have the representations
$$
(00000)\Qarrow(00010)\Qarrow(00020)\Qarrow\ldots\Qarrow(000n0)
\Qarrow\ldots\Eqn\TenSYMComplex
$$
The cohomology at zero momentum (the ``field content'')
can be calculated with purely
algebraic means, and is displayed in the table 1 below. Each
column represents the superfield content of the representations in the
complex (\TenSYMComplex), and has been shifted so that $Q$ acts
horizontally.

\vbox{
$$
\vtop{\baselineskip20pt\lineskip0pt
\ialign{
$\hfill#\quad$&$\,\hfill#\hfill\,$&$\,\hfill#\hfill\,$&$\,\hfill#\hfill\,$
&$\,\hfill#\hfill\,$&$\,\hfill#\hfill$&\qquad\qquad#\cr
            &n=0    &n=1    &n=2    &n=3    &n=4\cr
\hbox{dim}=0&(00000)&       &       &       &\phantom{(00000)}       \cr
        \fr2&\bullet&\bullet&               &       \cr
           1&\bullet&(10000)&\bullet&       &       \cr
       \Fr32&\bullet&(00001)&\bullet&\bullet&       \cr
           2&\bullet&\bullet&\bullet&\bullet&\bullet\cr
       \Fr52&\bullet&\bullet&(00010)&\bullet&\bullet\cr
           3&\bullet&\bullet&(10000)&\bullet&\bullet\cr
       \Fr72&\bullet&\bullet&\bullet&\bullet&\bullet\cr
           4&\bullet&\bullet&\bullet&(00000)&\bullet\cr
       \Fr92&\bullet&\bullet&\bullet&\bullet&\bullet\cr
}}
$$
\vskip2\parskip
\centerline{\it Table 1. The cohomology of the $D=10$ SYM complex.}
}
\vskip2\parskip
\noindent Our conventions for dimensions is that a bosonic derivative
has dimension 1, $[\partial_a]=1$. Then $[D_\a]=\fr2$, $[\l^\a]=-\fr2$
and $[Q]=0$.

The cohomology at non-zero momentum restricts the fields to obey the
linearised equations of motion. An easy way to read off the possible
equations of motion from the zero-momentum cohomology is to find the
cohomology in the column to the right of the one containing the
fields. The full non-linear super-Yang--Mills
equations of motion are reproduced by the solution at order $\l$ of
the zero curvature condition $Q\Psi+\Psi^2=0$. The equations may be
derived from a Chern--Simons like Batalin--Vilkovisky action where the
measure involves the position of the antighost (this requires
additional variables in order to make the measure non-degenerate
[\Berkovits], and will not be dealt with in this paper).

\section\DIsSix{Pure spinors and extended pure spinors in $D=6$}We
now turn our attention to $N=1$ super-Yang--Mills theory in $D=6$.
The conventions for Dynkin labels of $Spin(1,5)\times SU(2)$ used are
indicated in figure 2. The pure spinor $\l^\a$ is in the representation
$(001)(1)$ and $D_\a$ in $(010)(1)$.

We think of spinors
as 2-component quaternionic objects, with $\g^a$ (strictly speaking
$\sigma$-matrices, different for the two chiralities) acting from the
left and imaginary quaternions $e_i$, generating $SU(2)$, from the
right. Scalar product of two spinors $\a$ and $\b$ (of opposite
chiralities) is defined as $(\a\b)\equiv\Re(\a^\dagger\b)$, with
conjugation and real part being quaternionic.

\epsfxsize=3cm
\vskip2\parskip
\hskip8\parskip\vbox{\hsize=12cm
\epsffile{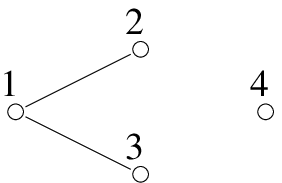}\hfill\break
{\it Figure 2: Convention for Dynkin indices for $so(1,5)\oplus su(2)$
representations}
}
\vskip2\parskip
If the procedure from the previous section is attempted,
a qualitatively different set of fields emerges in the cohomology.
The complex of representations now is
$$
(000)(0)\Qarrow(010)(1)\Qarrow(020)(2)\Qarrow\ldots\Qarrow(0n0)(n)
\Qarrow\ldots\Eqn\SixSYMComplexI
$$
and the zero-momentum cohomology is given in table 2.

\vskip12pt
\vbox{
$$
\vtop{\baselineskip20pt\lineskip0pt
\ialign{
$\hfill#\quad$&$\,\hfill#\hfill\,$&$\,\hfill#\hfill\,$&$\,\hfill#\hfill\,$
&$\,\hfill#\hfill\,$&$\,\hfill#\hfill$&\qquad\qquad#\cr
            &n=0    &n=1    &n=2    &n=3    &n=4\cr
\hbox{dim}=0&(000)(0)&&\phantom{(000)(0)}&\phantom{(000)(0)}
                                        &\phantom{(000)(0)}\cr
        \fr2&\bullet&\bullet&               &       \cr
           1&\bullet&(100)(0)&\bullet&       &       \cr
       \Fr32&\bullet&(001)(1)&\bullet&\bullet&       \cr
           2&\bullet&(000)(2)&\bullet&\bullet&\bullet\cr
       \Fr52&\bullet&\bullet&\bullet&\bullet&\bullet\cr
           3&\bullet&\bullet&\bullet&\bullet&\bullet\cr
       \Fr72&\bullet&\bullet&\bullet&\bullet&\bullet\cr
           4&\bullet&\bullet&\bullet&\bullet&\bullet\cr
       \Fr92&\bullet&\bullet&\bullet&\bullet&\bullet\cr
}}
$$
\vskip2\parskip
\centerline{\it Table 2. The cohomology of the $N=1$, $D=6$ SYM complex.}
}
\vskip2\parskip
The
condition $Q\Psi=0$ does not imply the equations of motion; instead
the $SU(2)$ triplet of auxiliary fields appears at dimension 2. There
is no cohomology at $n>1$, \ie, no antifields.
There are no representations for anti-fields, and consequently
no room for equations of motion, or currents.

The antifields turn out to be found in a separate complex,
$$
(000)(2)\Qarrow(010)(3)\Qarrow(020)(4)\Qarrow\ldots\Qarrow(0n0)(n+2)
\Qarrow\ldots\Eqn\SixSYMComplexII
$$
with the cohomology given by table 3.
Here, the dimension of the triplet field is chosen to be 2
in order to match with the dimensionalities of the current multiplet.

\vskip12pt
\vbox{
$$
\vtop{\baselineskip20pt\lineskip0pt
\ialign{
$\hfill#\quad$&$\,\hfill#\hfill\,$&$\,\hfill#\hfill\,$&$\,\hfill#\hfill\,$
&$\,\hfill#\hfill\,$&$\,\hfill#\hfill$&\qquad\qquad#\cr
            &n=0    &n=1    &n=2    &n=3    &n=4\cr
\hbox{dim}=2&(000)(2)&&\phantom{(000)(0)}&\phantom{(000)(0)}
                                &\phantom{(000)(0)}\cr
        \Fr52&(010)(1)&\bullet&               &       \cr
           3&(100)(0)&\bullet&\bullet&       &       \cr
       \Fr72&\bullet&\bullet&\bullet&\bullet&       \cr
           4&\bullet&(000)(0)&\bullet&\bullet&\bullet\cr
       \Fr92&\bullet&\bullet&\bullet&\bullet&\bullet\cr
           5&\bullet&\bullet&\bullet&\bullet&\bullet\cr
       \Fr{11}2&\bullet&\bullet&\bullet&\bullet&\bullet\cr
           6&\bullet&\bullet&\bullet&\bullet&\bullet\cr
       \Fr{13}2&\bullet&\bullet&\bullet&\bullet&\bullet\cr
}}
$$
\vskip2\parskip
\centerline{\it Table 3. The cohomology of the $N=1$, $D=6$ SYM
antifield complex.}
}
\vskip2\parskip

A key observation here is that both the fields and the antifields can
be accommodated in the same field $\Psi$, if it, in addition to the pure
spinor $\l$ of dimension $-\fr2$,
depends on a bosonic triplet $c_i$ of dimension $-2$
with $c_i(\l e_i)^\a=0$, where $e_i$, $i=1,2,3$, are imaginary
quaternions acting on the two-component $\l$ by right multiplication.
Then the $\l$ expansion at order $c$ gives exactly the
series of representations (\SixSYMComplexII). Multiplying this
condition one more time from the right with $c_ie_i$ gives the
necessary constraint $c_ic_i=0$.

The ``first quantised Hilbert space'' consists of functions of $\l^\a$,
$c_i$, $x^a$ and $\th^\a$, with the constraints $T^a\equiv(\bl\g^a\l)=0$,
$t^\a\equiv c_i(\l e_i)^\a=0$ and
$\tau\equiv c_ic_i=0$. Just like the constraint $T^a=0$ is solved by a
complexified $\l$, the full set of constraints are solved by complex
$\l$ and $c$. Any state is a formal series expansion in non-negative
powers
of $\l^\a$ and $c_i$. If we call this space $\PP$, the Hilbert space
is $\HH=\PP/T\PP$, where $T\PP$ is the ideal in $\PP$ generated by the
constraints.

$\l^\a$ is in (001)(1) and $c_i$ in (000)(2). The constraints imply
that $\l^nc^\nu$ only contains $(00n)(n+2\nu)$. We let $n$ and $\nu$
denote the degree of homogeneity in $\l$ and $c$, respectively,
throughout the paper.
The pure spinor BRST operator is $Q=(\bl D)$. The ghost and fields sit
in the cohomology of $Q$ at $c^0$ and the antifields and antighost at
$c^1$. The auxiliary field sits at $c^0\l^1\th^3$ and the corresponding
antifield at $c^1\l^0\th^0$, so one will need an operator $s\sim cwD^3$ to go
from one to the other. Here, $w_\a$ is the derivative with respect to $\l^\a$.
The derivative with respect to $c_i$ is denoted $b_i$.

There is also cohomology of $Q$ at higher powers of $c$, although
only at $\l^0$. The cohomologies at $c^\nu$ is given in Table 4, and
the general positions of non-vanishing cohomology in the power
expansion in $\l$ and $c$ is found in figure 3. We
are not sure what they signify, but note that they sit ``beyond'' the
scalar representation of the antifield in the power expansion, that
would provide a measure. Each of
these off-shell multiplets contain $8(2\nu-1)$ bosonic fields and the same
number of fermions.

\vskip6pt
\vbox{
$$
\vtop{\baselineskip20pt\lineskip0pt
\ialign{
$\hfill#\quad$&$\,\hfill#\hfill\,$&$\,\hfill#\hfill\,$&$\,\hfill#\hfill\,$
&$\,\hfill#\hfill\,$&$\,\hfill#\hfill$&\qquad\qquad#\cr
            &n=0    &n=1    &n=2    &n=3    &n=4\cr
\hbox{dim}=2\nu&(000)(2\nu)&\phantom{(000)(0)}&\phantom{(000)(0)}
                &\phantom{(000)(0)}&\phantom{(000)(0)}\cr
        2\nu+\fr2&(010)(2\nu-1)&\bullet&               &       \cr
           2\nu+1&(100)(2\nu-2)&\bullet&\bullet&       &       \cr
       2\nu+\Fr32&(001)(2\nu-1)&\bullet&\bullet&\bullet&       \cr
           2\nu+2&(000)(2\nu-4)&\bullet&\bullet&\bullet&\bullet\cr
       2\nu+\Fr52&\bullet&\bullet&\bullet&\bullet&\bullet\cr
           2\nu+3&\bullet&\bullet&\bullet&\bullet&\bullet\cr
       2\nu+\Fr72&\bullet&\bullet&\bullet&\bullet&\bullet\cr
           2\nu+4&\bullet&\bullet&\bullet&\bullet&\bullet\cr
       2\nu+\Fr92&\bullet&\bullet&\bullet&\bullet&\bullet\cr
}}
$$
\vskip2\parskip
\centerline{\it Table 4. The cohomology of $N=1$, $D=6$ at $\nu\geq2$.}
}

\font\llmath=cmsy18 at 36pt
\def\markthisbox{\raise5pt\hbox{\llmath\char'2}}

\vskip32pt
\hskip8\parskip\vbox{\hsize=12cm
\vbox{\tabskip=0pt\offinterlineskip
\halign{#&\hbox to 1.5cm{\hfil#}&#&\hbox to 1cm{\hfil#\hfil}&#&\hbox to 1cm{\hfil#\hfil}&#&\hbox to 1cm{\hfil#\hfil}&#&\hbox to 1cm{\hfil#\hfil}&#&\hbox to 1cm{\hfil#\hfil}&#&\hbox to 1cm{\hfil#\hfil}&#\cr
&\raise8pt\hbox{$n+\nu\,\,\,=$}&&\raise8pt\hbox{$0$}&&\raise8pt\hbox{$1$}&&\raise8pt\hbox{$2$}&&\raise8pt\hbox{$3$}&&\raise8pt\hbox{$4$}&&\raise8pt\hbox{$5$}&\cr
&&\vrule height.5pt&\vrule height.5pt width1cm&\vrule height.5pt&\vrule height.5pt width1cm&\vrule height.5pt&\vrule height.5pt width1cm&\vrule height.5pt&\vrule height.5pt width1cm&\vrule height.5pt&\vrule height.5pt width1cm&\vrule height.5pt&\vrule height.5pt width1cm&\vrule height.5pt\cr
&\raise10pt\hbox{$\nu=0\quad$}&\vrule height1cm&\markthisbox&\vrule&\markthisbox&\vrule&&\vrule&&\vrule&&\vrule&&\vrule\cr
&&\vrule height.5pt&\vrule height.5pt width1cm&\vrule height.5pt&\vrule height.5pt width1cm&\vrule height.5pt&\vrule height.5pt width1cm&\vrule height.5pt&\vrule height.5pt width1cm&\vrule height.5pt&\vrule height.5pt width1cm&\vrule height.5pt&\vrule height.5pt width1cm&\vrule height.5pt\cr
&\raise10pt\hbox{$1\quad$}&&&\vrule height1cm&\markthisbox&\vrule&\markthisbox&\vrule&&\vrule&&\vrule&&\vrule\cr
&&&&\vrule height.5pt&\vrule height.5pt width1cm&\vrule height.5pt&\vrule height.5pt width1cm&\vrule height.5pt&\vrule height.5pt width1cm&\vrule height.5pt&\vrule height.5pt width1cm&\vrule height.5pt&\vrule height.5pt width1cm&\vrule height.5pt\cr
&\raise10pt\hbox{$2\quad$}&&&&&\vrule height1cm&\markthisbox&\vrule&&\vrule&&\vrule&&\vrule\cr
&&&&&&\vrule height.5pt&\vrule height.5pt width1cm&\vrule height.5pt&\vrule height.5pt width1cm&\vrule height.5pt&\vrule height.5pt width1cm&\vrule height.5pt&\vrule height.5pt width1cm&\vrule height.5pt\cr
&\raise10pt\hbox{$3\quad$}&&&&&&&\vrule height1cm&\markthisbox&\vrule&&\vrule&&\vrule\cr
&&&&&&&&\vrule height.5pt&\vrule height.5pt width1cm&\vrule height.5pt&\vrule height.5pt width1cm&\vrule height.5pt&\vrule height.5pt width1cm&\vrule height.5pt\cr
&\raise10pt\hbox{$4\quad$}&&&&&&&&&\vrule height1cm&\markthisbox&\vrule&&\vrule\cr
&&&&&&&&&&\vrule height.5pt&\vrule height.5pt width1cm&\vrule height.5pt&\vrule height.5pt width1cm&\vrule height.5pt\cr
}}
\vskip2\parskip
\noindent{\it Figure 3. Position of cohomologies in the power
series.}
}
\vskip2\parskip

\section\NewOperator{The new BRST operator}Derivatives
are not well defined on $\HH$, since they can map an
element of the ideal to something outside the ideal. For example,
$w_\a\cdot(\bl\g^a\l)=2(\g^a\l)_\a\not\in T\PP$. One can define
modified ``quantum'' derivatives that map the ideal to itself, so that
they are well-defined on $\HH$. A good derivative, normalised so that it
acts exactly as $w$ on the representations $(0n0)(n+2\nu)$,
is \foot\dagger{In this and following expressions, no ``normal
ordering'' is assumed. Ordering is of course relevant, and all
expressions are ordered as they are written.}
$$
\tw_\a=\fr{2(1+n+2\nu)}\bigl[(2+n+2\nu)w_\a-j_i(w e_i)_\a\bigr]\komma
\eqn
$$
where $j_i$ is the $su(2)$ generator $j_i=(\bl we_i)+2\e_{ijk}c_jb_k$
and $n$ and $\nu$ are the number operators for $\l$ and $c$,
 $n=(\bl w)$, $\nu=c_ib_i$.
However, a careful analysis of the most general Ansatz for an operator $s$
shows that it cannot be constructed using such gauge invariant
operators only, if one demands that it must act in the cohomology of
$Q$. Instead, as a ``last resort'',
one may allow for an operator acting in the cohomology of $Q$ to be
not strictly gauge invariant, but gauge invariant modulo $Q$-exact
terms.

There is essentially two terms to write down. The Ansatz
is
$$
s=[Q,f(n)c_i(\bw\g^aw)(\bD\g_aDe_i)]+g(n)c_i(\bw\g^aDe_i)\partial_a
\komma\Eqn\SAnsatz
$$
where $f$ and $g$ are
some functions of $n$ (the functions could in principle depend also on $\nu$,
but everything else in $s$ commutes with $\nu$).
The first term contains $(\bw\g^aw)$, which is not gauge invariant,
but $[T^a,(\bw\g^bw)]=-4(j^{ab}+(n+4)\eta^{ab})$, where
$j^{ab}=(\bl\g^{ab}w)$, which is a gauge invariant operator (\ie,
well-defined on the ideal). Since $[Q,T^a]=0$, the first term in
eq. (\SAnsatz) is gauge invariant modulo $Q$-exact terms. An analogous
statement holds for invariance under $t^\a$.
The second term in eq. (\SAnsatz) is gauge invariant.

We will now check if it is possible to obtain $\{Q,s\}=0$ on $\HH$ for some
functions $f$, $g$. Let us write $s=s_1+s_2=[Q,u]+s_2$ for the terms above.
We then have
$$
\eqalign{
\{Q,s_1\}&=\{Q,[Q,u]\}=\fr2[\{Q,Q\},u]=-[T^a,u]\partial_a\cr
&=4c_if(n)(j^{ab}+(n+4)\eta^{ab})(\bD\g_bDe_i)\partial_a\punkt\cr
}\eqn
$$
Here, we have
dropped $[T^a,f(n)]=(f(n-2)-f(n))T^a$, which standing on the left
gives pure gauge (an element in the ideal $T\PP$). We also have
$$
[Q,s_2]=(g(n-1)-g(n))(\bl D)c_i(\bw\g^aDe_i)\partial_a
-g(n)c_i(\bD\g^aDe_i)\partial_a\punkt\eqn
$$
The term from the anticommutator of
the $D$'s in $Q$ and $s_2$ has been dropped; it vanishes thanks to
$t^\a=0$.
Now, we want to do a Fierz rearrangement of the first of these terms,
\ie, of $(\bl D)c_i(\bw\g^aDe_i)$. Doing a general Fierz in the two
$D$'s, one finds that only $(\bD\g_bDe_i)$ contributes (again, due to
$t^\a=0$), so that $(\bl D)c_i(\bw\g^aDe_i)
=-\fr4(j^{ab}-n\eta^{ab})c_i(\bD\g_bDe_i)$.
Taken together, this gives two types of terms in $[Q,s_1]$ and
$[Q,s_2]$, those with $j$ and those with $\eta$. Demanding that they
cancel gives the equations
$$
\eqalign{
j&:\quad 4f(n)+\fr4(g(n)-g(n-1))=0\komma\cr
\eta&:\quad 4(n+4)f(n)-g(n)-\Fr{n}4(g(n)-g(n-1))=0\punkt\cr
}\eqn
$$
The equations have a solution which is unique up to an overall normalisation:
$$
\eqalign{
f(n)&=-\Fr k{(n+2)(n+3)(n+4)}\komma\cr
g(n)&=\Fr{8k}{(n+3)(n+4)}\punkt\cr
}\eqn
$$

An explicit evaluation of the commutator $[Q,u]$ gives
$$
\eqalign{
s&=\Fr{3k}{(n+1)(n+2)(n+3)(n+4)}Qc_i(\bw\g^aw)(\bD\g_aDe_i)\cr
&-\Fr{2k}{(n+2)(n+3)(n+4)}c_i(\bw D^3_i)
+\Fr{8k}{(n+2)(n+3)}c_i(\bw\g^aDe_i)\partial_a\komma\cr
}\eqn
$$
where $(D^3)^\a_i=(\g^aD)^\a(D\g_a De_i)-8(\g^a De_i)^\a\*_a$ is the
totally antisymmetric product $\w^3 D$ in $(001)(3)$.
The first term vanishes when one chooses a gauge $(\bw\g^aw)\Psi=0$ for the
wave function, and vanishes in the cohomology of $Q$. The
remaining terms are the essential ones. It is obvious from the
structure of the cohomology of $Q$ (figure 3) that $s^2=0$ on any
element of the cohomology.

The calculations performed in this section make use of the properties
of antisymmetric tensor products of spinors, that dictate the content
of superfields. The upper half of a scalar superfield is shown in
figure 4 (the remaining part is of course given by the conjugate
representations).

\vskip12pt
\vskip2\parskip
\epsfysize=35mm
\hskip8\parskip\vbox{\hsize=12cm
\epsffile{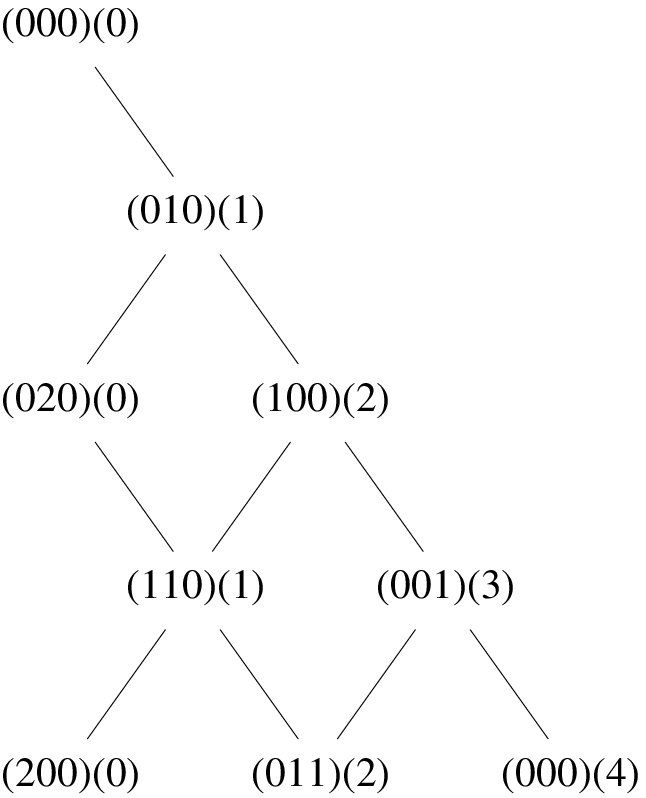}\hfill\break
{\it Figure 4: The antisymmetric products of chiral spinors in $D=6$}
}
\vskip2\parskip

\section\Conclusions{Conclusions and outlook}We
have enlarged the pure spinor space in $D=6$ to include also a
bosonic $SU(2)$ triplet $c_i$. The generalised pure spinor constraints
are $T^a\equiv(\bl\g^a\l)=0$,
$t^\a\equiv c_i(\l e_i)^\a=0$ and
$\tau\equiv c_ic_i=0$.
The cohomology of $Q=\l^\a D_\a$ in a scalar field then includes not
only ghosts and fields but also antifields and antighosts. We have
found a new BRST operator $s$ whose
cohomology imply the equations of motion. However, in order to define
$s$, we had to restrict to the cohomology of $Q$, outside of which $s$
is not well defined. This means that it is not possible to form a
modified BRST operator $Q+s$ on the scalar field. If one wants to
write a BV action for $D=6$, $N=1$ super-Yang--Mills (after
introducing some non-minimal set of pure spinor variables
[\Berkovits]),
it would need to be formulated on a field satisfying $Q\Psi+\Psi^2=0$.
This is clearly a weakness of the formalism, but it is not at all
clear that it is a generic one.

It will be interesting to continue the investigation for other
non-maximal supersymmetric theories. Supergravity with $N=1$
supersymmetry in $D=10$ is a natural candidate. Here, the
off-shell/on-shell discussion is more complicated, since the
superspace formulation contains some, but not all, equations of
motion. Preliminary results show that the structure is clearly
reflected in the corresponding pure spinor cohomology. There is also
indication that the situation may be better than for $D=6$, $N=1$
super-Yang--Mills theory, in the sense that it may be possible to
find a single modified BRST operator on the unconstrained field
depending on a generalised pure spinor [\CGN]. It is also
conceivable that a generalised pure spinor space provides a natural
setting for heterotic supergravity. Other interesting cases are
half-maximal euclidean supergravities in $D=6$ and $D=7$ in which
backgrounds for topological string theory and M-theory may be
embedded.

A superspace treatment of deformation of a non-maximally symmetric
theory, such as $D=6$ super-Yang--Mills, would rely on deformation of
the condition $s\Psi=0$. This would apply for Born--Infeld dynamics,
or its derivative expansion.

Another issue that may be addressed using similar methods
is the relation between the two superspace versions of (linearised)
11-dimensional supergravity. The same on-shell multiplet is found
using either a vector field (the ``vielbein complex'') or a scalar
field (the ``$C$-field complex''). Only the $C$-field complex is
appropriate for writing a linearised action. If one can enlarge the
pure spinor space to encompass the two complexes in the same field, it
may provide a starting point for the search for a superspace
action. Similar thoughts have been touched on in refs.
[\HoweTsimpis,\GrassiVanhove].

\acknowledgements We want to thank Niclas Wyllard for sharing his
expertise on pure spinors. MC would also
like to thank Michail Vasiliev, Dmitri
Sorokin and Per Sundell for discussions and constructive response to the
material in this paper.

\refout

\end